\documentclass[aps,prl,twocolumn,groupedaddress]{revtex4}

\usepackage{graphicx,color}
\usepackage{amsmath,amssymb}
\usepackage{amsbsy}

\newcommand{\up}{\uparrow}

\renewcommand{\d}{{\rm d}}

\newcommand{\imai}{{\rm i}}

\begin{document}
\title{Correlation-induced conductance suppression
at level degeneracy in a quantum dot}

\author{H.~A.~Nilsson}
\author{O.~Karlstr{\"o}m}
\author{M.~Larsson}
\author{P.~Caroff} 
\author{J.~N.~Pedersen}
\author{L.~Samuelson}
\author{A.~Wacker}
\author{L.-E.~Wernersson} 
\author{H.~Q.~Xu}\thanks{Corresponding author; hongqi.xu@ftf.lth.se}
\affiliation{Nanometer Structure Consortium, Lund University, Box 118, 221 00 Lund, Sweden}
\date{\today}

\begin{abstract}
The large, level-dependent g-factors in an InSb nanowire
quantum dot allow for the occurrence of a variety of level crossings in the dot.
While we observe the standard conductance enhancement in the Coulomb blockade
region for aligned levels with different spins due to the Kondo effect, a
vanishing of the conductance is found at the alignment of levels with equal
spins. This conductance suppression appears as a canyon cutting through the
web of direct tunneling lines and an enclosed Coulomb blockade region. In the
center of the Coulomb blockade region, we observe the predicted
correlation-induced resonance, which now turns out
to be part of a larger scenario. Our findings are supported by numerical and
analytical calculations.
\end{abstract}
\maketitle


Due to their small size the properties of nanosystems are dominated by
discrete quantum levels. These levels can be tuned by the means of, e.g.,
applying a magnetic field. At level crossings a variety of different physical
features can arise from electron coherence between the levels as well as
correlations with contacts. These can be conveniently probed by
transport measurements, see, e.g.,
Refs.~\cite{OosterkampPRL1998,ReimannRMP2002,PayettePRL2009}, where
generically conductance peaks can be associated with the presence of electronic
states at the Fermi level. While most
experimental work is done on GaAs-based systems where the individual levels
are approximately spin-degenerate due to the small electron g-factor, less
results exist at the crossing of spin-resolved levels. Such a level crossing
is particularly interesting if both levels are slightly below the Fermi
energy, so that there is one electron in the system while a second electron is
blocked to enter due to Coulomb repulsion. In this Coulomb blockade region the
conductance is typically small. At very low temperatures, 
however, the Kondo effect provides a strong conductance enhancement
at degeneracy of levels with opposite spins as observed
in Refs.~\cite{GoldhaberNature1998,CronenwettScience1998,SasakiNature2000}. 
For the case of degeneracy of levels with
equal spins, much less is known.
Here, recent theoretical calculations \cite{MedenPRL2006,KashcheyevsPRB2007,SilvestrovPRB2007}
showed a vanishing conductance at the point of electron-hole
symmetry in the middle of the Coulomb blockade region together with the
correlation-induced resonance, a strong
enhancement of the conductance for slight detuning between the levels.

In this letter, we present detailed measurements on InSb quantum dots realized
by electrically contacting epitaxially grown InSb nanowires. These devices
exhibit giant, strongly level-dependent electron g-factors
\cite{NilssonNL2009}, which allow for a clear observation of several
spin-resolved level crossings at relatively weak magnetic fields, and we can
directly compare the results of transport measurements at crossings of levels with equal and different spins.  For the case of
equal spins we are able to verify the predicted correlation-induced resonance
\cite{MedenPRL2006} in the center of the Coulomb blockade region. Furthermore
our data shows that this effect is a part of a larger scenario where
conductance suppression appears as a line in the parameter space of  detuning
and gate voltage. These findings are supported by numerical and analytical
calculations based on a two-level, equal-spin, interacting model, which fully
confirm the observed scenario.



\begin{figure}[t]
\includegraphics[width=0.9\columnwidth]{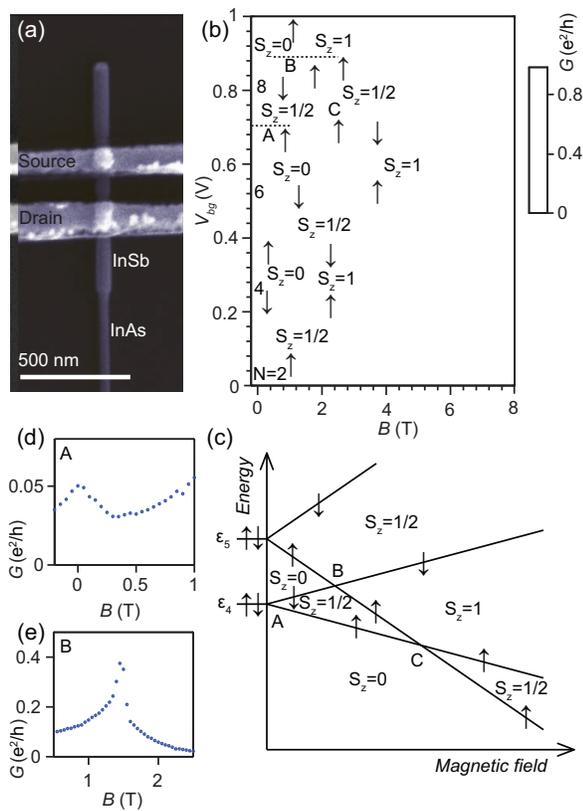}
\caption{ (Color online) 
  (a) SEM image of a fabricated InSb nanowire quantum-dot device. 
  (b) Conductance in grey scale measured at a source-drain bias of $V_{\rm
    sd}=0.5$ mV. (c) Schematic for the evolution of
  the single-particle levels 4 and 5 with magnetic field (neglecting the level
  interaction and Coulomb charging). Letters A, B, and C mark the three
  single-particle level
  degenerate points investigated in this work. (d) Conductance along line cut
  A in panel (b) showing a conventional spin-1/2 Kondo peak at $B=0$. 
  (e) Conductance along line cut B showing an integer-spin Kondo-like
  conductance enhancement at $B\approx 1.5$ T.
\label{FigSetupandOverview}}
\end{figure}

The InSb nanowire dot devices investigated here are fabricated from InSb
segments of InAs/InSb heterostructure nanowires where the InAs segments are
used as seed nanowires to favor nucleation of InSb 
\cite{NilssonNL2009,CaroffSmall2008,SuppMat}.
Figure~\ref{FigSetupandOverview}(a) shows a scanning electron
microscope (SEM) image of a fabricated device, where the dot is formed
between two 150-nm-wide Ti/Au contacts with a distance of 100 nm in an 
InSb nanowire with a
diameter of 65 nm. All measurements are performed in a $^3$He cryostat 
at 300 mK.


Figure~\ref{FigSetupandOverview}(b) shows a grey-scale plot of the
conductance, $G$, as a function of the magnetic field $B$ and the back-gate
voltage $V_{\rm bg}$ applied to the Si substrate of a fabricated device. 
The spin state of the last filled electron
in each energy level is indicated by an arrow in the figure. Here negative
values of the g-factor are assumed for all quantum levels
\cite{IsaacsonPR1968}. From the magnetic field evolutions of the conductance
peaks, we can evaluate the electron g-factors for the dot levels
(following Refs.~\cite{RoddaroPRL2008,LarssonAPL2009}) and find that the values of
these g-factors are giant, with the largest absolute value reaching about 60,
and are strongly level-dependent (see also Refs.~\cite{NilssonNL2009,SuppMat}). A
large difference between the electron g-factors in the dot allows for the crossings of the
spin-up state of the 5th level with both the spin-down and the spin-up state
of the 4th level as $B$ is increased. Figure~\ref{FigSetupandOverview}(c)
shows a schematic for the scenario of such single-particle level crossings
without taking level interaction and the effect of Coulomb charging into account.


The transport measurements in Fig.~\ref{FigSetupandOverview}(b) display
several clear signatures of both the conventional spin-1/2 Kondo and the
integer-spin Kondo-like effects which manifest themselves as conductance
enhancements in the Coulomb blockade regions in the InSb quantum dot. For
example a weak but
visible conductance ridge is observed at zero magnetic field inside the $N = 7$
Coulomb blockade region. A line plot of the conductance as a function of the
magnetic field along cut A
is shown in
Fig.~\ref{FigSetupandOverview}(d). Here one can easily identify a conductance
peak at zero magnetic field. This peak occurs at the standard spin-degeneracy
point of level 4 of the quantum dot at $B=0$ as indicated by label A in the
schematic shown in Fig.~\ref{FigSetupandOverview}(c). Similar conductance peaks
or ridges in odd-number electron Coulomb blockade regions are observed in
several other fabricated InSb quantum dots at zero magnetic field. All of
these conductance enhancements can be attributed to the conventional spin-1/2
Kondo effect \cite{GoldhaberNature1998,CronenwettScience1998}.

In addition to the spin-1/2 Kondo effect we 
see clear signatures of
integer-spin Kondo-like correlations
\cite{SasakiNature2000,NygardNature2000}. One such example is provided by the
clear high-conductance ridge in the $N = 8$ Coulomb blockade region at $B\sim
1.5$ T [see Fig.~\ref{FigSetupandOverview}(b)]. A corresponding line plot of
the conductance as a function of the magnetic field along 
the cut B through the
conductance ridge is shown in Fig.~\ref{FigSetupandOverview}(e), where a
conductance peak is clearly observed. This peak occurs at the degeneracy of
levels 4 and 5 with opposite spins as indicated by label B in the schematic
shown in Fig.~\ref{FigSetupandOverview}(c). An additional, though weaker,
integer-spin Kondo-like conductance enhancement is
observed in the $N =
6$ Coulomb blockade region at $B\approx3$ T. 
Such integer-spin Kondo-like conductance enhancements appear at the transition 
from a spin singlet to the $S_z=1$ state of an $S=1$ spin triplet
as the magnetic field increases. Note that the other states of the
$S=1$ triplet are significantly higher in energy due to the large
g-factors. Thus more complex phenomena such as the two-stage Kondo
effect \cite{WielPRL2002,PustilnikPRL2001} are not observed here. Both the
spin-1/2 and the integer spin Kondo-like {\em enhancements} of the conductance
in the Coulomb blockade region occur when there are {\em two degenerate states 
of different spins}.

\begin{figure}
\includegraphics[width=0.9\columnwidth]{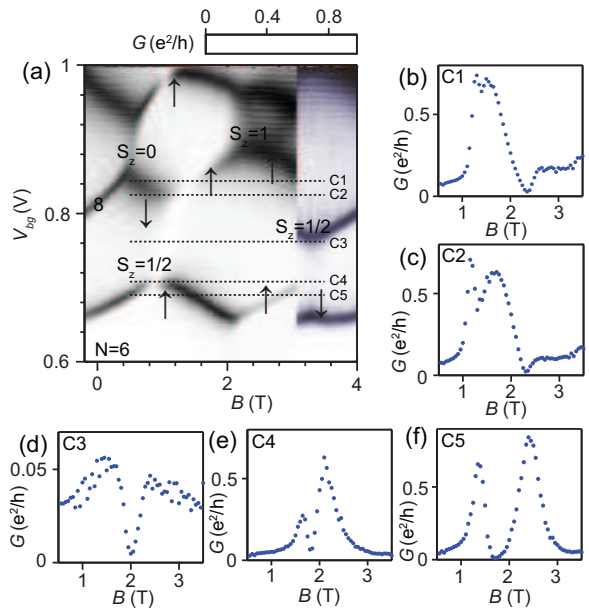}
\caption{(Color online)
Details of the conductance suppression.
(a)  Enlarged section of Fig.~\ref{FigSetupandOverview}(b).
(b)-(f) Conductance plots along line cuts C1-C5 
(at back-gate voltages of 848, 822, 762, 708, and 690 mV) 
of panel (a), respectively.
\label{FigExpCanyon}}
\end{figure}

 
We now focus our attention to the $N = 7$ Coulomb blockade region at magnetic
fields of $B\sim 2$ T [Fig.~\ref{FigSetupandOverview}(b)] where the 4th and
the 5th level with the {\em same spin} cross as indicated by label C in the
schematic shown in Fig.~\ref{FigSetupandOverview}(c). Here we observe a clear
{\em suppression of the conductance}  within the cotunneling background in the
Coulomb blockade region. Moreover the direct tunneling lines are also broken
at the crossing points at the corners of the Coulomb blockade region. This
scenario is shown in detail in Fig.~\ref{FigExpCanyon}(a).
The bright region
of the conductance suppression resembles a canyon which connects the upper $S_z=1$ and
$N=8$ Coulomb blockade region with the lower $S_z=0$ and $N=6$  region while
cutting through both the direct tunneling lines and the $S_z=1/2$ and $N=7$
blockade region. As shown in Figs.~\ref{FigExpCanyon}(b)-\ref{FigExpCanyon}(f) 
for different gate
voltages, the conductance drops approximately down to zero at the bottom of the
canyon. In addition, the conductance in the middle of the Coulomb blockade region
[Fig.~\ref{FigExpCanyon}(d)] shows a clear enhancement on
both sides of the conductance suppression. This is the
correlation-induced resonance, which was 
predicted for a similar, strong correlated quantum dot system near the electron-hole symmetry point \cite{MedenPRL2006}.

This canyon of conductance suppression is the main finding of our
letter. We note that the presence of
giant, strongly level-dependent g-factors in our InSb nanowire dot is crucial
to create the degeneracy between the two spin-up levels at a moderate magnetic
field.  Furthermore, the quadratic shifts of the levels with magnetic field
differ strongly for levels 4 and 5, which has helped to create this desired
degeneracy point.  
Intuitively, the phenomenon of the conductance suppression can be understood as a result of 
(i) the strong modification of the dot states by correlations
with the contacts and (ii) the consequent (destructive) 
interference between the two paths through the dot 
associated with the two modified states. This is straightforwardly seen
in the lower and upper direct tunneling regions, see the Breit-Wigner results below.
However, in the Coulomb
blockade region, strong correlations between contacts and the dots states
exist, which can result in vanishing conductance at the
electron-hole symmetry point at zero temperature and zero bias
\cite{MedenPRL2006,KashcheyevsPRB2007,SilvestrovPRB2007}.
The observed canyon of conductance suppression connects both scenarios
and suggests that the combination of correlations and interference is required
for a full understanding.


In the following we show theoretically that a crossing between 
two levels with equal spins indeed provides a canyon of conductance suppression
that cuts through both
the Coulomb blockade and direct tunneling regions.
Our Hamiltonian $\hat{H}_D + \hat{H}_C$ combines the terms 
describing the quantum dot and its coupling
to the left (L) and right (R) lead which read 
(similar to 
Refs.~\cite{MedenPRL2006,KashcheyevsPRB2007,SilvestrovPRB2007})
\begin{eqnarray}
\hat{H}_D&=&E_{4\up}a^\dag_{4\up}a_{4\up}+E_{5\up}a^\dag_{5\up}a_{5\up}+
Ua^\dag_{4\up}a_{4\up}a^\dag_{5\up}a_{5\up}\, ,\label{EqHD}\\
\hat{H}_C&=&\sum_{k,\ell=L/R} t_\ell(k)c_{\ell\up}(k)
(a^\dag_{4\up}+x_\ell a^\dag_{5\up})+\textrm{h.c.}\nonumber \\
&&+\sum_{k,\ell=L/R}E_{\ell}(k)c^\dag_{\ell\up}(k)c_{\ell\up}(k)\label{EqHC}\, ,
\end{eqnarray}
where $a_{i\up}\, (a_{i\up}^{\dag})$ and $c_{\ell\up}\, (c_{\ell\up}^{\dag})$ are the annihilation (creation)
operators of electrons in the dot and leads, respectively.
We define $\Gamma_{\ell 4}(E) = 2\pi\sum_k |t_\ell(k)|^2
\delta[E_{\ell}(k)-E]$, which is assumed to be constant, and $\Gamma_{\ell
  5}=|x_\ell|^2\Gamma_{\ell 4}$. 
Fitting to the conductance peaks \cite{SuppMat},
we obtain $U=5\; \mathrm{ meV}$, $\Gamma_{L4} = 0.3$ meV,
$\Gamma_{R4}=0.1$ meV, $\Gamma_{L 5}=1$ meV,  $\Gamma_{R5}= 0.4$ meV, and
the level energies of $E_{4\up} = 2\; 
\mathrm{ meV}\times (B/\mathrm{T} - 2) - E_{\rm g}-U/2$ and  $E_{5\up} = -2.5\; 
\mathrm{ meV}\times(B/\mathrm{T}-2)- E_{g}-U/2$.
Here, the gate level energy $E_{g}$ corresponds to the back gate voltage by $E_{g}=
e(V_{bg}-767\; \mathrm{ mV})/22$. 
(This choice of $E_g$ provides electron-hole symmetry around $B=2$ T at
$E_{g}=0$.) As motivated below, we use $x_L<0$  and
$x_R>0$, which might reflect a parity difference between the fourth and fifth quantum orbital states.
The occupations of the left and right
leads are given by Fermi functions with the electrochemical potentials of $\pm eV_{\rm sd}/2$,
respectively.

\begin{figure}
\includegraphics[width=1.0\columnwidth]{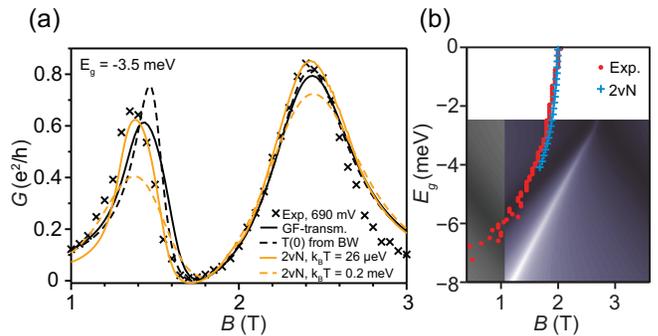}
\caption{(Color online) Conductance where interaction is of minor importance.
(a) Experimental data (crosses) for $V_{\rm bg}=690$ mV together with calculated 
results using the noninteracting Green's function (GF) model with
$V_{\rm sd}=0.5$ mV (black solid line) and 
the Breit-Wigner (BW) transmission formula of Eq.~(\ref{EqBW}) (black dashed
line). Corresponding 2vN results, including interactions, 
are given in thin orange lines for two temperatures.
(b) Conductance from the GF model in its region of validity, $E_g \lesssim -U/2$,
together with the positions of the minimal conductance  from the experiment 
(red dots) and the 2vN model (blue crosses).
\label{FigCondNoInt}
}
\end{figure}

For $E_g\lesssim -U/2$, 
at most one of the two levels is occupied and Coulomb interaction plays no
role. In such
a non-interacting system, the transmission $T(E)$ can be calculated with Green's
functions (GF), see, e.g., Ref.~\cite{ShahbazyanPRB1994} where a similar system
was treated.  The finite bias conductance reads  
$G=(e/hV_{\rm sd})\int_{-eV_{\rm sd}/2}^{eV_{\rm sd}/2}\d E\, T(E)$  
for zero
temperature, see the black solid
line in Fig.~\ref{FigCondNoInt}(a), which 
agrees well with the data (crosses) from Fig.~\ref{FigExpCanyon}(f).
Figure~\ref{FigCondNoInt}(b) shows that the experimental canyon of conductance
suppression is reproduced very well by the GF model in its range of validity,
$E_g\lesssim -U/2$. 
Here the vanishing conductance can be attributed to
interference between the transmission through both levels.
The Breit-Wigner formula provides
\begin{multline}
T(E)=\Gamma_{L4}\Gamma_{R4}\Big|\frac{1}{E-E_{4\up}+\imai 
(\Gamma_{L4}+\Gamma_{R4})/2}\\
+\frac{x_Lx_R}{E-E_{5\up}+\imai 
(\Gamma_{L5}+\Gamma_{R5})/2}\Big|^2\label{EqBW}\, .
\end{multline}
We  find the vanishing of
$T(0)$ at $x_Lx_RE_{4\up}\approx -E_{5\up}$ assuming 
$\Gamma_{Li}+\Gamma_{Ri}\ll E_{i\up}$. For our parameters this provides $E_g+ U/2\approx
3.7\; \mathrm{ meV}\times(B/\mathrm{T}-2)$ 
in good agreement with the numerical findings and the experimental data 
shown in Fig.~\ref{FigCondNoInt}(b). This
justifies our choice of $x_Lx_R<0$.

\begin{figure}
\includegraphics[width=0.9\columnwidth]{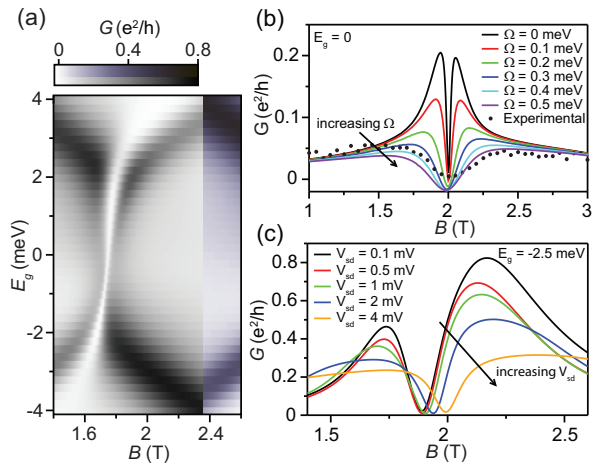}
\caption{(Color online) Calculations by the 2vN approach at $k_BT=0.2$ meV, $V_{\rm sd}=0.5$ mV
and $\Omega=0$ unless stated otherwise.
(a) Canyon of conductance suppression.
(b) Conductance for different interlevel couplings 
$\Omega$ together with the experimental data.  (c) Conductance for different bias values of $V_{\rm sd}$.
\label{Fig2vN}
}
\end{figure}

In the Coulomb blockade region ($-U/2<E_g<U/2$) 
the current is carried by cotunneling events. 
Here we apply the second-order von Neumann (2vN)  approach
\cite{PedersenPRB2005a}, which treats all interactions in $\hat{H}_D$ exactly.
Correlated transitions between the leads and the dot states
are included in second order describing cotunneling \cite{PedersenPHE2010}
and interference \cite{PedersenPRB2009}.
For $E_g=-3.5$ meV, the results agree excellently
with the experimental data, see Fig.~\ref{FigCondNoInt}(a). The 2vN is only
reliable above $T_K$ \cite{PedersenPRB2005a} which we estimate to be
$k_BT_K\approx 0.1$ meV \cite{SuppMat} and we attribute the occurrence of some slightly
negative conductivities to the improper treatment of higher-order
processes. In order to reduce these problems we use the increased temperature
of $k_BT=0.2$ meV in the  following. The results of the
calculation are given in Fig.~\ref{Fig2vN}(a) and show the canyon of 
conductance suppression in good agreement with the experiment.

The calculated conductance 
shows an approximate symmetry around the
electron-hole symmetry point (full symmetry is restored by
reversing the bias).  
In contrast, the experimental data are
more asymmetric, which indicates a gate voltage dependence of system 
parameters. Figure~\ref{Fig2vN}(b) 
shows that the conductance suppression persists
for a finite interlevel coupling described by an additional term
$\Omega a^\dag_{5\up}a_{4\up} +h.c.$ to $\hat{H}_D$
in Eq.~(\ref{EqHD}), while its width increases.
Comparing with experimental data, $\Omega\approx 0.4$ meV fits better
at $V_\mathrm{bg}=762$ mV.
Even higher values of $\Omega$ seem appropriate for larger $V_\mathrm{bg}$ (not
shown), while $\Omega\approx 0$ fits well for $V_\mathrm{bg}=690$ mV (see Fig.~\ref{FigCondNoInt}). 
This indicates that the interlevel coupling depends on the back-gate
voltage in our dot and vanishes accidentally around $V_\mathrm{bg}\approx
700$ mV. Finally,
Fig.~\ref{Fig2vN}(c) shows that the vanishing of the current persists for higher
biases. In the high-bias limit with infinite $U$  this corresponds to the
situation discussed in Ref.~\cite{LiEPL2009}, where it was shown that the
current vanishes exactly at level degeneracy (i.e., at $B=2$ in our case), independently of the
couplings unless $x_L=x_R$.

In {\em conclusion} we have observed that the crossing of quantum levels with
equal spins in the presence of Coulomb repulsion manifests as a canyon of
vanishing conductance cutting through the direct tunneling lines and the
enclosed Coulomb blockade region. This scenario is well covered by the 2vN
approach based on a two-level, equal-spin, interacting model. 
Furthermore, our experimental data confirms the predicted
correlation-induced resonances close to the electron-hole symmetry point.
Our results show that a full understanding of the interplay
between strong correlations and interference is required to describe the
entire behavior of the conductance of the system at degeneracy of levels with equal spins.

\acknowledgments We thank Karsten Flensberg and Feng Zhai for stimulating
discussions and Claes Thelander for technical help. This work was supported by the Swedish Research
Council (VR), the Swedish Foundation for Strategic Research (SSF), and the Knut and Alice Wallenberg Foundation.


\end{document}